\documentclass[12pt]{article}
\begin{document}

\vspace*{1.0 cm}

\begin{center}
{\Large \bf Toward an Infinite-component Field Theory with a Double Symmetry:
Interaction of Fields} \\

\vspace{1.0 cm}
\renewcommand{\thefootnote}{*}
{\large L.M.Slad\footnote{E-mail: slad@theory.sinp.msu.ru}} \\

\vspace{0.4 cm}

{\it Skobeltsyn Institute of Nuclear Physics, \\
Moscow State University, Moscow 119899}
\end{center}

\vspace{0.5 cm}

\centerline{\bf Abstract}

\begin{small}
We complete the first stage of constructing a theory of fields not investigated
before; these fields transform according to Lorentz group representations
decomposable into an infinite direct sum of finite-dimensional irreducible
representations. We consider only those theories that initially have a double
symmetry: relativistic invariance and the invariance under the transformations
of a secondary symmetry generated by the polar or the axial four-vector
representation of the orthochronous Lorentz group. The high symmetry of the
theory results in an infinite degeneracy of the particle mass spectrum with
respect to spin. To eliminate this degeneracy, we postulate a spontaneous
secondary-symmetry breaking and then solve the problems on the existence and
the structure of nontrivial interaction Lagrangians.
\end{small}

\vspace{0.5 cm}

\begin{center}
{\large \bf 1. Introduction}
\end{center}

We recently began constructing a theory of fields that transform according 
to representations of the proper Lorentz group $L^{\uparrow}_{+}$ that are
decomposable into an infinite direct sum of finite-dimensional irreducible
representations (we call such fields the ISFIR-class fields) [1]. This
construction is based on a group theory approach formulated in [2] and called
a double symmetry. It aims to study a possible of effectively describing 
hadrons in terms of the monolocal infinite-component fields.
  
A special feature of the general relativistically invariant theory of the free
ISFIR-class fields is an infinite number of arbitrary constants in its
Lagrangians of the form
\begin{equation}
{\cal L}_{0} = \frac{i}{2}[(\Psi, \Gamma^{\mu} \partial_{\mu} \Psi ) - 
(\partial_{\mu} \Psi, \Gamma^{\mu} \Psi )] - (\Psi, R\Psi )
\end{equation}
and in their corresponding Gelfand--Yaglom equations [3], [4]. Such constants
correspond to each pair of irreducible $L^{\uparrow}_{+}$-group representations
such that the matrix elements of the four-vector operator $\Gamma^{\mu}$ are
nonzero. Because of the infinite arbitrariness with respect to constants,
nobody analyzed the infinite-component ISFIR-class field theory or spoke about
its physical properties until [1]. The elemination of this arbitrariness was 
the main problem posed in [1] and solved there by extracting that part of the
general ISFIR-class field theory having relativistic invariance (the primary
symmetry) that is additionally invariant under global secondary-symmetry
transformations of the form
\begin{equation}
\Psi (x) \rightarrow \Psi '(x) = \exp [-i D^{\mu} \theta_{\mu}] \Psi (x),
\end{equation}
where $D^{\mu}$ are matrix operators and the parameters $\theta_{\mu}$ are the
components of the polar or the axial four-vector of the orthochronous Lorentz
group $L^{\uparrow}$.

The notion of the secondary symmetry introduced in [2] generalizes the
long-known symmetries like the symmetry of the Gell-Mann--Levy $\sigma$-model
[5] and supersymmetry (see, e.g., [6]). By definition, the secondary-symmetry
transformations do not break the primary symmetry, and their parameters belong
to the space of a certain representation of the primary-symmetry group. In the
$\sigma$-model, the parameters of secondary-symmetry transformations are the
pseudoscalars with respect to the orthochronous Lorentz group, whereas
supersymmetry is generated by the bispinor representation of the proper 
Lorentz group. The secondary symmetry used in [1], whose transformations are
given by relation (2), is a natural member of this series of the topical
symmetries of field theories. Transformations (2) connect sectors belonging to
the spaces of different irreducible $L^{\uparrow}_{+}$-group representations 
because of which the problem of eliminating the infinite arbitrariness in the
constants of the relativistically invariant theory of the ISFIR-class fields
can be completely solved.

In each of the variants of the free infinite-component field theory with the
double symmetry that were found and described in [1], we know, first, a
representation of the proper Lorentz group, according to which the field
transforms, second, the four-vector operator $D^{\mu}$ from secondary-symmetry
transformations (2) (up to a common numerical factor), and, third, the
four-vector operator $\Gamma^{\mu}$ and therefore the kinetic term of
Lagrangian (1) (up to a common numerical factor). It is also known that the
operator $R$ defining the mass term of Lagrangian (1) is a multiple of the
identity operator.

In complete accordance with the known Coleman--Mandula theorem [7] relating to
consequences of the Lorentz-group extension, the mass spectrum corresponding
to any of the variants of the infinite-component field theory with the double
symmetry is infinitely degenerate with respect to spin. This requires
introducing a spontaneous secondary-symmetry breaking that leaves the primary
symmetry (the relativistic invariance) of the theory unchanged. To be specific,
we suppose that the scalar (with respect to the group $L^{\uparrow}$)
components of one or several bosonic infinite-component ISFIR-class fields have
nonzero vacuum expectation values. A spontaneous secondary-symmetry breaking 
can introduce only one correction into the free-field theory with the double 
symmetry described in [1], namely, it can change the operators $R$, i.e.,
change the mass terms in the Lagrangians of form (1). To find those variants of
the theory for which such nonzero changes are possible and to obtain new
expressions for the operators $R$, we must solve the problem of the existence
and the structure of nontrivial interaction Lagrangians for the
infinite-component fields with the initial double symmetry.

In this paper, we confine ourselves to finding only those Lagrangians knowing
which completes the construction of the free fermionic ISFIR-class field
theories and brings us right up to the beginning of the study of a number of
physical properties of such fields: their mass spectra and the characteristics
of electromagnetic interaction. Constructing the free and the self-interacting
bosonic ISFIR-class field theories should be completed only in the case where
describing baryons in terms of the considered infinite-component fields proves
successful.

We now present the subject matter of our paper in more concrete formulations.

We analyze the structure of three-particle interaction Lagrangians of 
the form\footnote[1]{All the necessary information about the notions used and
notation relating to the Lorentz group can be found in [1].}  
\begin{equation}
{\cal L}_{\rm int} = \sum_{i,\tau,l,m} 
\overline{\psi} (x) Q^{\tau lm}_{i} \psi (x) \varphi_{\tau lm}^{i} (x)
\equiv \sum_{i,\tau ,l,m} 
(\psi (x), Q^{\tau lm}_{i} \varphi_{\tau lm}^{i} (x) \psi (x)),
\end{equation}
where $\psi (x)$ is a fermion field, $\varphi_{\tau lm}^{i} (x)$ is the
bosonic-field component characterized by the irreducible
$L^{\uparrow}_{+}$-group representation $\tau = (l_{0}, l_{1})$, by the spin 
$l$, and by its third-axis projection $m$, and  $Q^{\tau lm}_{i} \equiv 
Q^{(l_{0}, l_{1})lm}_{i}$ are matrix operators. The index $i$ indicates that
according to conditions 1 and 3A in [1], two types of bosonic field that differ
by the transformation properties under space reflection are included in
Lagrangian (3).

We consider only those Lagrangians (3) for which the bosonic fields have the
scalar or the pseudoscalar (with respect to the group $L^{\uparrow}$) 
components (these fields are respectively denoted by $\varphi^{+}(x)$ and
$\varphi^{-}(x)$, and the index $i$ is assigned the values $+$ and $-$).
Therefore, at this stage of constructing the theory, only two representations,
$S^{1}$ given by (I.88) and $S^{B}$ given by (I.89)\footnote[2]{Here and in
what folows, the references to the formulas in [1] are given in the form (I.N),
where N is the number of the corresponding formula.}, are selected from the
countable set of the proper Lorentz group representations with integer spin
that was found in [1]. 
 
We impose the requirement on the interaction Lagrangians, as well as on the
free Lagrangians in [1], that they initially have the double symmetry:
relativistic invariance and the invariance under secondary-symmetry
transformations (2) or (I.94) generated by the polar or the axial four-vector
representation of the orthochronous Lorentz group.

If the scalar (with respect to the group $L^{\uparrow}$) bosonic-field 
component $\varphi^{+}_{(0,1)00}(x)$ acquires a nonzero vacuum expectation
value $\lambda$ as a result of a spontaneous secondary-symmetry breaking, then
the sum of Lagrangians (I.12) and (3) results in fermionic-field Lagrangian (1)
with the operator $R$ of the form
\begin{equation}
R = \kappa E + \lambda Q^{(0,1)00}_{+},
\end{equation}
where $E$ is the identity operator. The operator $Q^{(0,1)00}_{+}$ depends on
the ratio of the normalization constants for the four-vector operators 
$D^{\mu}$ in the secondary-symmetry transformations of the bosonic and the
fermionic fields, this ratio plays the role of a parameter. At the stage of
comparing the properties of the infinite-component fields and the properties of
hadrons, we consider a sum of Lagrangians of type (3) that contain several 
bosonic fields that differ, in particular, in the mentioned normalization 
constants. In this case, formula (4) contains several terms of the type 
$\lambda Q^{(0,1)00}_{+}$. 

The subject matter is further presented in the following order. First, we write
formulas involving the four-vector operators; these formulas are repeatedly 
used to solve the problems in the paper. Next, we successively consider the 
question about the existense of the nontrivial fermion-boson interaction 
Lagrangians and the matrix elements of the operators $Q^{(0,1)00}_{\pm}$ for 
the cases where the bosonic fields are described by the respective 
$L^{\uparrow}_{+}$-group representations $S^{1}$ and $S^{B}$. In conclusion, 
we give an example of a fermionic infinite-component ISFIR-class field with
interesting mass spectrum properties described by the constructed
double-symmetry theory. We note that the proofs of the formulas given in this
paper are based on evident algebraic operations and, as a rule, require lengthy
calculations.

\begin{center}
{\large \bf 2. Some relations involving four-vector operators}
\end{center}

We take two four-vector operators $V^{\mu}$ and $W^{\mu}$ defined by the
respective quantities $v_{\tau' \tau}$ and $w_{\tau' \tau}$. From the analogues
of formulas (I.17)--(I.23), as well as from relations (I.7)--(I.10), we obtain
\begin{equation}
\sum_{l',m'} g_{\mu\nu} V^{\mu}_{\tau'' l''m'', \tau' l'm'}
W^{\nu}_{\tau' l'm', \tau lm} = b_{\tau \tau'} v_{\tau \tau'} w_{\tau' \tau}
\delta_{\tau'' \tau} \delta_{l''l} \delta_{m''m},
\end{equation}
where 
\begin{equation}
b_{\tau \tau'} = \left\{
\begin{array}{ll}
2(l_{1}-l_{0}-1)(l_{1}+l_{0}+1), & {\mbox{\rm if}} \hspace{0.3cm} 
\tau'=(l_{0}+1,l_{1}), \\
2(l_{1}-l_{0}+1)(l_{1}+l_{0}-1), & {\mbox{\rm if}} \hspace{0.3cm}
\tau'=(l_{0}-1,l_{1}), \\
2(l_{0}-l_{1}-1)(l_{1}+l_{0}+1), & {\mbox{\rm if}} \hspace{0.3cm}
\tau'=(l_{0},l_{1}+1), \\
2(l_{0}-l_{1}+1)(l_{1}+l_{0}-1), & {\mbox{\rm if}} \hspace{0.3cm}
\tau'=(l_{0},l_{1}-1),
\end{array} \right.
\end{equation}
for $\tau = (l_{0},l_{1})$. There is no summation over the repeated index 
$\tau'$  in the left side of relation (5).

In the free-field theory with the double symmetry, the operators $D^{\mu}$ in
transformations (2) satisfy the relation 
\begin{equation}
D^{\mu}D_{\mu} =HE,
\end{equation}
where $H$ is a numerical factor. This relation is a consequence (I.26) and is
easily verified by direct calculations.

Using formulas (5) and (6) and the values of the quantities $d_{\tau'\tau}$
chracterizing the four-vector operator $D^{\mu}$ and written in Corollaries
1--4 in [1], we have
\begin{equation}
H = \left\{
\begin{array}{rl}
4d^{2}_{0} & {\mbox{\rm for Corollaries 1 and 3 (point 1),}} \\
-4d^{2}_{0} & {\mbox{\rm for Corollary 2 (point 2),}} \\
-8d^{2}_{0} & {\mbox{\rm for Corollary 2 (point 3),}} \\
8d^{2}_{0} & {\mbox{\rm for Corollary 3 (point 2),}}
\end{array} \right.
\end{equation}
where $d_{0}=g_{0}c_{0}$.

We now introduce the four-vector operator $\Omega^{\mu}$ described by the
quantities $\omega_{\tau' \tau}$ and constructed from the given operators
$V^{\mu}$ and $W^{\mu}$ as
$$\Omega^{\mu} = V^{\nu}W^{\mu}V_{\nu}.$$
Using the analogues of formulas (I.17)--(I.23) and relations (I.7)--(I.10), we
obtain
$$\frac{1}{2} \omega (l_{0}+1,l_{1};l_{0},l_{1}) =$$
$$= (l_{1}-l_{0}+1)(l_{1}+l_{0}-1) v(l_{0}+1,l_{1};l_{0},l_{1})
w(l_{0},l_{1};l_{0}-1,l_{1}) v(l_{0}-1,l_{1};l_{0},l_{1})$$
$$- v(l_{0}+1,l_{1};l_{0},l_{1}) w(l_{0},l_{1};l_{0}+1,l_{1})
v(l_{0}+1,l_{1};l_{0},l_{1})$$
$$+ (l_{1}-l_{0}-2)(l_{1}+l_{0}+2) v(l_{0}+1,l_{1};l_{0}+2,l_{1})
w(l_{0}+2,l_{1};l_{0}+1,l_{1}) v(l_{0}+1,l_{1};l_{0},l_{1})$$
$$+ (l_{1}+l_{0}-1) v(l_{0}+1,l_{1};l_{0},l_{1})
w(l_{0},l_{1};l_{0},l_{1}-1) v(l_{0},l_{1}-1;l_{0},l_{1})$$
$$- (l_{1}-l_{0}-2)(l_{1}+l_{0}-1) v(l_{0}+1,l_{1};l_{0}+1,l_{1}-1)
w(l_{0}+1,l_{1}-1;l_{0},l_{1}-1) v(l_{0},l_{1}-1;l_{0},l_{1})$$
$$- (l_{1}-l_{0}+1) v(l_{0}+1,l_{1};l_{0},l_{1})
w(l_{0},l_{1};l_{0},l_{1}+1) v(l_{0},l_{1}+1;l_{0},l_{1})$$
$$- (l_{1}-l_{0}+1)(l_{1}+l_{0}+2) v(l_{0}+1,l_{1};l_{0}+1,l_{1}+1)
w(l_{0}+1,l_{1}+1;l_{0},l_{1}+1) v(l_{0},l_{1}+1;l_{0},l_{1})$$
$$+ (l_{1}-l_{0}-2) v(l_{0}+1,l_{1};l_{0}+1,l_{1}-1)
w(l_{0}+1,l_{1}-1;l_{0}+1,l_{1}) v(l_{0}+1,l_{1};l_{0},l_{1})$$
\begin{equation}
- (l_{1}+l_{0}+2) v(l_{0}+1,l_{1};l_{0}+1,l_{1}+1)
w(l_{0}+1,l_{1}+1;l_{0}+1,l_{1}) v(l_{0}+1,l_{1};l_{0},l_{1}).
\end{equation}

The expression for $\frac{1}{2} \omega (l_{0},l_{1};l_{0}+1,l_{1})$ is obtained
from (9) by replacing $v_{\tau' \tau}$, $w_{\tau' \tau}$, and 
$\omega_{\tau' \tau}$ with $v_{\tau \tau'}$, $w_{\tau \tau'}$ and 
$\omega_{\tau \tau'}$, and the formulas for $\frac{1}{2} 
\omega (l_{0},l_{1}+1;l_{0},l_{1})$ and $\omega (l_{0},l_{1};l_{0},l_{1}+1)$ 
are obtained from the respective formulas for $\frac{1}{2} 
\omega (l_{0}+1,l_{1};l_{0},l_{1})$ and $\frac{1}{2} 
\omega (l_{0},l_{1};l_{0}+1,l_{1})$ by replacing $l_{0} \leftrightarrow l_{1}$. 

If the four-vector operator $D^{\mu}$ from secondary-symmetry transformations
(2) corresponds to the variants of the free-field theory described in Corollary
2 (point 3) or in Corollary 3 (point 2) in [1], then using formula (9), we
obtain  
\begin{equation}
D^{\nu}D^{\mu}D_{\nu}=0.
\end{equation}
For the same operators $D^{\mu}$, the relation 
\begin{equation}
\varepsilon_{\mu\nu\rho\sigma}D^{\mu}D^{\nu}D^{\rho}D^{\sigma}=0
\end{equation}
also holds.

It is evident that interaction Lagrangian (3) can be invariant under
transformations (2) or (I.94), if and only if their generated 
secondary-symmetry group is common to both the fermionic and the bosonic 
fields. All the secondary-symmetry groups obtained in [1] can be referred to 
one of the following three types.

The first type includes the Abelian groups, for which
\begin{equation}
[D^{\mu}, D^{\nu}]=0.
\end{equation}
This equality holds for the operators $D^{\mu}$ corresponding to Corollaries 1,
2 (point 2), and 3 (point 1) in [1]. For instance, formulas (I.75) and (I.76)
relating to Corollaries 1 and 3 (point 1) imply that $D^{\mu} = 
g_{0}\Gamma^{\mu}$, and relation (I.26) then leads to equality (12). If the
operators $D^{\mu}$ are described by formulas (I.79)--(I.82) in Corollary 2
(point 2), then we can verify the relations that are equivalent to equality
(12) and are obtained from Eqs. (I.32)--(I.44) by replacing $c_{\tau' \tau} 
\rightarrow d_{\tau' \tau}$.

The second type of the secondary-symmetry groups corresponds to Corollaries 2
(point 3) and 3 (point 2) in [1]. For this type, the antisymmetric operators
$G^{\mu\nu}$ defined by
\begin{equation}
G^{\mu\nu} = [D^{\mu}, D^{\nu}]
\end{equation}
are nonzero. The equalities
\begin{equation}
[G^{\mu\nu}, G^{\rho\sigma}] = 0,
\end{equation}
\begin{equation}
\{ D^{\mu}, G^{\nu\rho} \} = 0
\end{equation}
hold. Using relation (15), we can show that the secondary-symmetry groups of 
the second type are infinite.

The third type of the secondary-symmetry groups corresponds to Corollaries 2
(point 1) and 4 in [1]. For this type, the inequality 
$[G^{\mu\nu}, G^{\rho\sigma}] \neq 0$ holds.

Until now, we used the same notion for a number of quantities defined in the
spaces of both the fermionic and the bosonic fields. In what follows, we
include the additional indices $F$ and $B$ in the notation of such quantities:
$D^{F\mu}$, $H^{F}$, and $G^{F\mu\nu}$ and $D^{B\mu}$, $H^{B}$, and 
$G^{B\mu\nu}$. 

We now list the variants of interaction Lagrangians (3) to be considered in
accordance with the group properties of the fields involved.

{\bf Variants 1 and 1A}. Each of the bosonic fields transforms according to the
$L^{\uparrow}_{+}$-group representation $S^{1}$ given by (I.88). The fermionic
field trasforms according to any of the representations $S^{k_{1}}$ given by 
(I.72). The secondary-symmetry transformations are generated by the polar and 
the axial four-vector representations of the group $L^{\uparrow}$ in the 
respective variants 1 and 1A. Such bosonic-field transformations have form (2) 
and (I.94) in the respective variants 1 and 1A. The operators $D^{B\mu}$ are 
described by formulas (I.74) and (I.76), where $k_{1}=1$. The 
secondary-symmetry transformations for the fermionic field have form (2). The 
operators $D^{F\mu}$ involved are described by formulas (I.73)--(I.76) and 
(I.79)--(I.82) in the respective variants 1 and 1A.

{\bf Variant 2}. Each of the bosonic and the fermionic fields transforms 
according to the respective $L^{\uparrow}_{+}$-group representation
$S^{B}$ given by (I.89) and $S^{F}$ given by (I.83). The secondary-symmetry 
transformations for the bosonic and the fermionic fields have the respective
forms (I.94) and (2), and the parameter $\theta_{\mu}$ in formulas (I.94) and
(2) is the axial four-vector of the group $L^{\uparrow}$. The operators
$D^{B\mu}$ and $D^{F\mu}$ are described by the respective formulas
(I.90)--(I.93) and (I.84)--(I.87).

The secondary-symmetry groups for variants 1 and 1A of the interaction
Lagrangian refer to the first type and for variant 2 do to the second type.

In what follows, we present detailed calculations relating to the structure of
the interaction Lagrangians only for variants 1 and 2; for variant 1A, we
indicate the changes that must be introduced in variant 1.

\begin{center}
{\large \bf 3. The representation $S^{1}$ for the bosonic field and the
fermion-boson interaction Lagrangian}
\end{center}

We let $Q_{\pm}$ denote the scalar (with respect to the group 
$L^{\uparrow}_{+}$) operator $Q^{(0,1)00}_{\pm}$ in Lagrangian (3). Let the
fermionic field $\psi (x)$ belong to the representation space $S$ of the
orthochronous Lorentz group, and let its Lorentz transformation have form 
(I.3). Then the fulfillment of the relation
\begin{equation}
S^{-1}(g) Q_{\pm} S(g) = Q_{\pm}
\end{equation}
is the necessary condition for the relativistic invariance of Lagrangian (3).
Here and in what follows, either the upper or the lower signs are taken
simultaneously.

From relation (16), we find that for any irreducible $L^{\uparrow}_{+}$-qroup
representation $(l_{0},l_{1})$ belonging to the representation $S$, the
equaities 
\begin{equation}
Q_{\pm} \xi_{(l_{0},l_{1})lm} = q^{\pm}(l_{0},l_{1}) \xi_{(l_{0},l_{1})lm},
\end{equation}
\begin{equation}
q^{\pm}(-l_{0},l_{1}) = \pm q^{\pm}(l_{0},l_{1}),
\end{equation}
where the quantities $q(|l_{0}|,l_{1})$ are arbitrary, must hold.

It is easy to prove that the fulfillment of relation (16) and equalities (21)
given below and expressing the operators $Q^{\tau lm}_{\pm}$ in Lagrangian (3) 
in terms of the operators $Q_{\pm}$ is the enough condition for the
relativistic invariance of Lagrangian (3).

Taking equality (I.27) into account, we can easily see that secondary-symmetry
transformations (2) generated by the polar four-vector of the group
$L^{\uparrow}$ leave Lagrangian (3) unchanged if the operators $Q^{\tau lm}$
satisfy the system of equations
\begin{equation}
[D^{F \mu}, Q^{\tau lm}_{\pm}] =
\sum_{\tau',l',m'} Q^{\tau' l'm'}_{\pm} D^{B \mu}_{\tau' l'm', \tau lm},
\end{equation}
where $\tau, \tau' \in S$ and $D^{B \mu}_{\tau' l'm', \tau lm}$ is the matrix
element of the operator $D^{B \mu}$.

Equations (19) with formulas (5) and (6) give a possibility to express the
operators $Q^{\tau_{1} lm}_{\pm}$ and $Q^{\tau_{0} lm}_{\pm}$ in terms of each
other for any preassigned $\tau_{1}$ and $\tau_{0}$ belonging to the
representation $S$, and this is possible in an infinite number of ways. The
main question is therefore about the equivalence of all of such expressions or,
what is the same, about the consistency of system of equations (19).

The following statement holds.  

{\bf Proposition 1.} {\it Let the transformation properties of the fields in
Lagrangian} (3) {\it coincide with those listed in variant} 1. {\it Then system 
of equations} (19) {\it is equivalent to the equations} 
\begin{equation}
D^{F \mu} Q_{\pm} D^{F}_{\mu}
= (H^{F} -H^{B}/2) Q_{\pm}
\end{equation}
{\it in the operator} $Q_{\pm}$ {\it and the system of independent equalities}
$$Q^{(0,n+1)lm}_{\pm} = (n+1) \left( \frac{2}{H^{B}} \right)^{n} \times$$
\begin{equation}
\times [D^{F}_{\nu_{n}}, \ldots, [D^{F}_{\nu_{2}},[D^{F}_{\nu_{1}},Q_{\pm}]] 
\ldots ] \left( D^{B\nu_{1}}D^{B\nu_{2}}\ldots D^{B\nu_{n}} 
\right)_{(0,1)00,(0,n+1)lm},
\end{equation}
{\it where} $n \geq 1$ {\it and the quantities} $H^{F}$ {\it and} $H^{B}$ 
{\it and given by respective relations} (7) {\it and} (8). 

{\bf Proof.} We first prove that Eqs. (20) and equalities (21) are the
consequence of system of equations (19). 

We multiply both sides of relation (19) by the operator $D^{F \nu}$ first from
the left an then from the right. We subtract the second of the obtained
equations from the first one and eiminate the operator $D^{F \nu}$ in the right
side of the resulting equation using initial Eq. (19). We have
\begin{equation}
[D^{F \nu},[D^{F \mu},Q^{\tau lm}_{\pm}]] =
\sum_{\tau',l',m'} Q^{\tau' l'm'}_{\pm} 
\left( D^{B \nu}D^{B \mu} \right)_{\tau' l'm', \tau lm}.
\end{equation}
We multiply both sides of this equation by $g_{\mu\nu}$ and sum over the
indices $\mu$ and $\nu$. Taking relation (7) into account and setting
$\tau = (0,1)$, $l=0$, and $m=0$, we obtain Eqs. (20).

We now take Eq. (19) with $\tau = (0,j)$ and $j \geq 1$. We multiply both sides
by $g_{\mu\nu}D^{B \nu}_{(0,j)lm, (0,j+1)l''m''}$ and sum over the indices 
$\mu$, $\nu$, $l$, and $m$. Using relatios (5), (6), (I.74), (I.76), and (8),
we obtain
$$Q^{(0,j+1)l''m''}_{\pm} =\frac{j+1}{j}\cdot\frac{2}{H^{B}}
\sum_{lm} [D^{F}_{\nu},Q^{(0,j)lm}_{\pm}] D^{B \nu}_{(0,j)lm, (0,j+1)l''m''},$$
whence equalities (21) immediately follow.

We now prove that all equations of system (19) are fulfilled if equalities
(21) and Eqs. (20) are fulfilled. 

We first verify that the relation 
$$\sum_{lm} \left( D^{B\nu_{1}}D^{B\nu_{2}} \ldots D^{B\nu_{n}}
\right)_{(0,1)00,(0,n+1)lm} \left( D^{B\mu_{n}} \ldots 
D^{B\mu_{2}}D^{B\mu_{1}} \right)_{(0,n+1)lm,(0,1)00}=$$
\begin{equation}
= s_{0}(n) \sum g^{\nu_{1} \mu_{j_{1}} } g^{\nu_{2} \mu_{j_{2}} }
\ldots g^{\nu_{n} \mu_{j_{n}} }
+ s_{1}(n) \sum g^{\nu_{i_{1}} \nu_{i_{2}} } g^{\mu_{j_{1}} \mu_{j_{2}} }
g^{\nu_{i_{3}} \mu_{j_{3}} }...g^{\nu_{i_{n}} \mu_{j_{n}} } + \ldots 
\end{equation}
holds. The first sum in the right side of this equality is taken over all
permutations of the indices $\mu_{1}, \mu_{2}, \ldots , \mu_{n}$. The second
sum is taken over all combinations of two indices $\nu_{i_{1}}$ and 
$\nu_{i_{2}}$ extracted from the collection $\nu_{1},\nu_{2},\ldots , \nu_{n}$,
over all combinations of two indices $\mu_{j_{1}}$ and $\mu_{j_{2}}$ extracted
from the collection $\mu_{1}, \mu_{2}, \ldots , \mu_{n}$, and over the
permutations of all remaining indices $\mu_{j}$ with some fixed order of the
remaining indices $\nu_{i}$ and so on.

Let $a^{i}_{\nu_{i}}$ and $b^{j}_{\mu_{j}}$, where $i,j=1,\ldots, n$, be
the covariant components of arbitrary four-vectors. We multiply the left side 
of formula (23) by $a^{1}_{\nu_{1}}\ldots a^{n}_{\nu_{n}}b^{n}_{\mu_{n}}\ldots 
b^{1}_{\mu_{1}}$ and take the sum over $\nu_{1},\ldots, \nu_{n},\mu_{n},\ldots, 
\mu_{1}$. The obtained expression ${\cal A}$ is invariant under the proper
Lorentz group transformations. According to the first main theorem of the
invariant theory [8], the expression ${\cal A}$ is a function of the standard
basic invariants for the group $L^{\uparrow}_{+}$ that are the scalar product 
of two four-vectors $(p^{1}p^{2}) \equiv g^{\mu\nu}p^{1}_{\mu}p^{2}_{\nu}$ and
the component determinant of four four-vectors $[p^{1}p^{2}p^{3}p^{4}] \equiv 
\varepsilon^{\mu\nu\rho\sigma} p^{1}_{\mu}p^{2}_{\nu}p^{3}_{\rho}
p^{4}_{\sigma}$. Because different components of the operator $D^{B\mu}$
commute with each other in the considered variant of Lagrangian (3), the
expression ${\cal A}$ remains unchanged under a permutation of any two vectors 
$a^{i}_{\nu_{i}}$ and $a^{i'}_{\nu_{i'}}$ or any two vectors $b^{j}_{\mu_{j}}$ 
and $b^{j'}_{\mu_{j'}}$. Consequently, the expression ${\cal A}$ cannot contain
the component determinants. Taking the arbitrariness of the auxiliary
four-vectors $a^{i}_{\nu_{i}}$ and $b^{j}_{\mu_{j}}$ into account, we thus
verify the validity of relation (23).

We mutipy both sides of equality (23) by $g_{\nu_{1}\nu_{2}}$ and sum over the
indices $\nu_{1}$ and $\nu_{2}$. As a result, the left side of the obtained
equality vanishes according to relation (5). We thus have
\begin{equation}
s_{1}(n) = - s_{0}(n)/n.
\end{equation}
We multiply both sides of equality (23) by $g_{\nu_{n}\mu_{n}}$ and sum over
indices $\nu_{n}$ and $\mu_{n}$. Using formulas (5), (6), (I.74), (I.76), and
(8), we obtain
$$(n+3)s_{0}(n) + (n-1)s_{1}(n)=\frac{n+1}{n} \cdot \frac{H^{B}}{2} 
s_{0}(n-1).$$
Using relation (24) and the equality $s_{0}(1)=H^{B}/4$, we thus obtain
\begin{equation}
s_{0}(n)= \frac{1}{(n+1)!} \left( \frac{H^{B}}{2} \right)^{n}.
\end{equation}

We now multiply both sides of equality (21) by $(D^{B\mu_{n}} \ldots 
D^{B\mu_{2}}D^{B\mu_{1}})_{(0,n+1)lm,(0,1)00}$ and use relations (23)--(25),
commutativity (12) of different components of the operator $D^{F\mu}$, formula
(7), and condition (20), which the operators $Q_{\pm}$ satisfy. We have
$$[D^{F \mu_{n}}, \ldots, [D^{F \mu_{2}},[ D^{F \mu_{1}},Q_{\pm}]] \ldots ]$$
$$- \frac{H^{B}}{2n} \sum g^{\mu_{i} \mu_{j}}
[D^{F \mu_{n}}, \ldots, [D^{F \mu_{2}},[ D^{F \mu_{1}},Q_{\pm}]] \ldots 
]_{\mu_{i}\mu_{j}}+\ldots =$$
\begin{equation}
= \sum_{l,m} Q^{(0,n+1)lm}_{\pm}\left( D^{B\mu_{n}} \ldots D^{B\mu_{2}} 
D^{B\mu_{1}}\right)_{(0,n+1)lm,(0,1)00}.
\end{equation}
Here and in what follows, $[D^{F \mu_{n}}, \ldots, [D^{F \mu_{1}},Q_{\pm}] 
\ldots ]_{\mu_{i_{1}} \ldots \mu_{i_{k}}}$ is the $(n-k)$-fold commutator
obtained from the $n$-fold commutator $[D^{F \mu_{n}}, \ldots, [D^{F \mu_{1}},
Q_{\pm}] \ldots ]$ by deleting the operators $ D^{F \mu_{i_{1}}}, \ldots, 
D^{F \mu_{i_{k}}}$. The first sum in the left side of equality (26) is taken
over all combinations of two indices $\mu_{i}$ and $\mu_{j}$ extracted from the
collection $\mu_{1}, \mu_{2}, \ldots , \mu_{n}$, and so on.

We also need one more formula obtained from relations (5), (6), (I.74), (I.76),
and (8), namely,
$$(D^{B\nu_{n}} \ldots D^{B\nu_{2}}D^{B\nu_{1}})_{(0,n+1)l'm',(0,1)00}
(D^{B}_{\nu_{1}}D^{B}_{\nu_{2}}\ldots D^{B}_{\nu_{n}})_{(0,1)00,(0,n+1)lm}=$$
\begin{equation}
=\frac{1}{n+1} \left( \frac{H^{B}}{2} \right)^{n} \delta_{l l'} \delta_{m m'}.
\end{equation}

Successively using equalities (21), (26), (7), (12), (27), and again (21), we
now obtain the chain of equalities
$$[D^{F \mu}, Q^{(0,n+1)lm}_{\pm}] =$$
$$= (n+1) \left( \frac{2}{H^{B}} \right)^{n}
[D^{F \mu},[D^{F}_{\nu_{n}}, \ldots, [D^{F}_{\nu_{1}},Q_{\pm}]\ldots ]]\left( 
D^{B\nu_{1}}\ldots D^{B\nu_{n}} \right)_{(0,1)00,(0,n+1)lm} =$$
$$= \left\{ (n+1)\left( \frac{2}{H^{B}} \right)^{n} \sum_{l'm'} 
Q^{(0,n+2)l'm'}_{\pm} \left( D^{B \mu} D^{B}_{\nu_{n}} \ldots  D^{B}_{\nu_{1}}
\right)_{(0,n+2)l'm',(0,1)00} \right.$$
$$+ \left. \left( \frac{2}{H^{B}} \right)^{n-1}
\sum_{i=1}^{n} \delta^{\mu}_{\nu_{i}}
[D^{F}_{\nu_{n}}, \ldots, [ D^{F}_{\nu_{1}},Q_{\pm}]\ldots ]_{\nu_{i}}\right\}
\left( D^{B\nu_{1}}\ldots D^{B\nu_{n}} \right)_{(0,1)00,(0,n+1)lm} =$$
$$= \sum_{l',m'} \left( Q^{(0,n+2)l'm'}_{\pm} D^{B \mu}_{(0,n+2)l'm',(0,n+1)lm}
+ Q^{(0,n)l'm'}_{\pm} D^{B \mu}_{(0,n)l'm',(0,n+1)lm} \right),$$
which ultimately reproduces the equalities of system (19) corresponding to
variant 1 of Lagrangian (3). We note that those terms in formulas (23) and (26)
that are not written explicity and a part of the terms in the sum before the
dots contribute zero to the final result of the chain of equalities written
above. This is due to the appearing of expressions $g_{\nu_{i}\nu_{j}}
(D^{B\nu_{1}} \ldots D^{B\nu_{n}})_{(0,1)00,(0,n+1)lm}$ whose factors reduce to
the form $\sum_{l'm'} g_{\mu\nu}D^{B\mu}_{(0,1)00,(0,2)l'm'}$ 
$D^{B\nu}_{(0,2)l'm',(0,3)l''m''}$ because of the commutativity of 
different components of the operator $D^{B\mu}$ given by (12). The vanishing of
the last sum is ensured by relation (5). The proof of Proposition 1 is thus 
completed.

We write relation (20) as a system of equations in the quantities 
$q(l_{0},l_{1})$ defined by formula (17). Using formulas (5)--(8) and
(I.73)--(I.76), we have
$$(k_{1}-l_{0}-1)(k_{1}+l_{0})q^{\pm}(l_{0}+1,l_{1})
+ (k_{1}-l_{0})(k_{1}+l_{0}-1)q^{\pm}(l_{0}-1,l_{1})$$
$$- (k_{1}-l_{1}-1)(k_{1}+l_{1})q^{\pm}(l_{0},l_{1}+1)
- (k_{1}-l_{1})(k_{1}+l_{1}-1)q^{\pm}(l_{0},l_{1}-1) =$$
\begin{equation}
= z(l_{1}-l_{0})(l_{1}+l_{0})q^{\pm}(l_{0},l_{1}),
\end{equation}
where $z = 2-H^{B}/H^{F}$ for all irreducible $L^{\uparrow}_{+}$-group 
representations $(l_{0},l_{1})$ belonging to representation $S^{k_{1}}$ given 
by (I.72) ($k_{1}$ is a half-integer, $k_{1} \geq 3/2$). In the variant under
consideration, we have $H^{B} > 0$ and $H^{F} > 0$; therefore, 
$z \in (-\infty, 2)$.

It is easy to see that nontrivial solutions of system of equations (18) and
(28) exist and the general solution contains $k_{1}-1/2$ arbitrary constants, 
for which we can take the quantities  $q(l_{0},k_{1})$ with $1/2 \leq l_{0} 
\leq k_{1}-1$. Directly using Eqs. (18) and (28) as recursive relations can be
preferable to using the explicit form of the quantities $q(l_{0},l_{1})$. Even
in the simplest case, where $k_{1}= 3/2$, the quantities $q(l_{0},l_{1})$ are
expressed in terms of higher transcendental functions,
\begin{equation}
q^{\pm}\left( \frac{1}{2},\frac{3}{2}\right) = q_{0}^{\pm}, \hspace{0.4 cm}
q^{\pm}\left( \frac{1}{2},l_{1}\right) = \frac{2q_{0}^{\pm}}{l_{1}^{2}-1/4}
\left[ C^{2}_{l_{1}-3/2} \left( \frac{z}{2} \right) \mp C^{2}_{l_{1}-5/2} 
\left( \frac{z}{2} \right) \right], \hspace{0.4 cm} l_{1} \geq \frac{5}{2},
\end{equation}
where $q_{0}^{\pm}$ are arbitrary constants and $C^{\nu}_{n}(x)$ is the
Gegenbauer polynomial [9]. Expressing the Gegenbauer polynomial in terms of the
hypergeometric series $F(a,b;c;y)$, we can represent formulas (29) as
$$q^{\pm}\left( \frac{1}{2},2N-\frac{1}{2}\right) =  q_{0}^{\pm} 
\frac{(-1)^{N+1}}{2N-1} \left[ F\left( 1-N,N+1; \frac{1}{2}; 
\frac{z^{2}}{4}\right) \right.$$
\begin{equation}
\left. \pm (N-1)zF\left( 2-N,N+1; \frac{3}{2}; \frac{z^{2}}{4}\right)\right],
\end{equation}
$$q^{\pm}\left( \frac{1}{2},2N+\frac{1}{2}\right) = q_{0}^{\pm} 
\frac{(-1)^{N}}{2N+1} \left[ \pm F\left(1-N,N+1; \frac{1}{2}; 
\frac{z^{2}}{4}\right) \right.$$
\begin{equation}
\left. - (N+1)z F\left( 1-N,N+2; \frac{3}{2}; \frac{z^{2}}{4}\right)\right],
\end{equation}
where $N \geq 1$.

When passing from variant 1 of Lagrangian (3) to variant 1A, we replace the
indices $\pm$ with the indices $\mp$ in the right side of Eqs. (19) and, if $n$ 
is odd, also in the left side of Eqs. (21) and in the right side of relations
(26). In the chain of equalities, we need to make such a replacement in the
third and fourth links and, if $n$ is odd, also in the second link. The system
of equations for the quantities $q^{\pm}(l_{0},l_{1})$ are identical in 
variants 1 and 1A. According to relation (10), we have $H^{B} > 0$ and 
$H^{F} < 0$ in variant 1A; therefore, $z \in (2, +\infty)$.

In variants 1 and 1A, each representation $S^{k_{1}}$ ($k_{1}$ is a
half-integer, $k_{1} \geq 3/2$) for the fermionic field is thus assigned
nontrivial Lagrangian (3) containing $k_{1}-1/2$ arbitrary constants. The 
matrix elements of the operators $Q_{\pm}$ are found from Eqs. (18) and (28).
The operators $Q^{(0,n+1)lm}_{\pm}$ are expressed in terms of the operators
$Q_{\pm}$ using equalities of type (21).

\begin{center}
{\large \bf 4. The representation $S^{B}$ for the bosonic field and the
fermion-boson interaction Lagrangian}
\end{center}

{\bf Proposition 2.} {\it Let the transformation properties of the fields in 
Lagrangian} (3) {\it coincide with the ones listed for variant} 2. {\it Then a
nontrivial Lagrangian} (3) {\it does not exist}.

{\bf Proof.} The invariance of Lagrangian (3) under the secondary-symmetry
transformations is provided by the fulfillment of conditions (19), in whose
right side it is necessary to replace the indices $\pm$ with $\mp$. These
conditions imply, first, Eqs. (20) for the operators $Q_{\pm}$ and, second, the
relations
\begin{equation}
[G^{F \mu\nu}, Q^{\tau lm}_{\pm}] =
\sum_{\tau' l'm'} Q^{\tau' l'm'}_{\pm} 
G^{B \mu\nu}_{\tau' l'm', \tau lm}.
\end{equation}
From relations (32), we can obtain the analogue of Eqs. (22), in which the
four-vector operators are replaced by the operators $G^{\mu\nu}$ and 
$G^{\rho\sigma}$. We multiply the obtained analogue of Eqs. (22) first by 
$g_{\mu\rho}g_{\nu\sigma}$ and then by $\varepsilon_{\mu\nu\rho\sigma}$. In
both cases, we then take the sum over the indices $\mu, \nu, \rho$, and 
$\sigma$ and take formulas (13), (7), (10), and (11) into account. Then two
further pairs of equations,
\begin{equation}
G^{F \mu\nu}Q_{\pm}G^{F}_{\mu\nu}=[(H^{B})^{2}-2(H^{F})^{2}]Q_{\pm},
\end{equation}
\begin{equation}
\varepsilon_{\mu\nu\rho\sigma}G^{F \mu\nu}Q_{\pm}G^{F\rho\sigma}=0,
\end{equation}
are added to Eqs. (20) for the operator $Q_{\pm}$.

Using formulas (5), (6), (8)--(10), and (I.84)--(I.87), Eqs. (20), (33), and
(34) are reduced to a system of equations for the quantities 
$q^{\pm}(l_{0},l_{1})$ in which 
$(l_{0},l_{1}) \in S^{F}$, and the representation $S^{F}$ is given by formula 
(I.83). If $l_{0}+l_{1}$ is an even number, we have
$$(l_{1}-l_{0}-1)q^{\pm}(l_{0}+1,l_{1}) + (l_{1}-l_{0}+1)q^{\pm}(l_{0}-1,l_{1})
+ (l_{1}-l_{0}+1)q^{\pm}(l_{0},l_{1}+1)$$
\begin{equation}
+ (l_{1}-l_{0}-1)q^{\pm}(l_{0},l_{1}-1)
= 2(2-y)(l_{1}-l_{0})q^{\pm}(l_{0},l_{1}),
\end{equation}
$$\frac{l_{1}+l_{0}-2}{l_{1}+l_{0}-1} q^{\pm}(l_{0}-1,l_{1}-1)
+\frac{l_{1}+l_{0}+2}{l_{1}+l_{0}+1} q^{\pm}(l_{0}+1,l_{1}+1)$$
\begin{equation}
+\frac{2}{(l_{1}+l_{0}-1)(l_{1}+l_{0}+1)} q^{\pm}(l_{0},l_{1})
=(2-y^{2})q^{\pm}(l_{0},l_{1}),
\end{equation}
$$(l_{1}-l_{0}+1)q^{\pm}(l_{0}-1,l_{1}+1)
+(l_{1}-l_{0}-1)q^{\pm}(l_{0}+1,l_{1}-1)=$$
\begin{equation}
= (2-y^{2})(l_{1}-l_{0}) q^{\pm}(l_{0},l_{1});
\end{equation}
if $l_{0}+l_{1}$ is an odd number, then
$$(l_{1}+l_{0}+1)q^{\pm}(l_{0}+1,l_{1})
+ (l_{1}+l_{0}-1)q^{\pm}(l_{0}-1,l_{1})
+ (l_{1}+l_{0}+1)q^{\pm}(l_{0},l_{1}+1)$$
\begin{equation}
+ (l_{1}+l_{0}-1)q^{\pm}(l_{0},l_{1}-1) 
= 2(2-y)(l_{1}+l_{0})q^{\pm} (l_{0},l_{1}),
\end{equation}
$$(l_{1}+l_{0}-1)q^{\pm}(l_{0}-1,l_{1}-1)
+(l_{1}+l_{0}+1)q^{\pm}(l_{0}+1,l_{1}+1)=$$
\begin{equation}
= (2-y^{2})(l_{1}+l_{0}) q^{\pm}(l_{0},l_{1}),
\end{equation}
$$\frac{l_{1}-l_{0}+2}{l_{1}-l_{0}+1} q^{\pm}(l_{0}-1,l_{1}+1)
+\frac{l_{1}-l_{0}-2}{l_{1}-l_{0}-1} q^{\pm}(l_{0}+1,l_{1}-1)$$
\begin{equation}
+\frac{2}{(l_{1}-l_{0}-1)(l_{1}-l_{0}+1)} q^{\pm}(l_{0},l_{1})
=(2-y^{2})q^{\pm}(l_{0},l_{1}),
\end{equation}
where $y=H^{B}/H^{F}$. In variant 2, the inequalities $H^{B} > 0$, $H^{F} < 0$,
and $y < 0$ hold according to relation (10).

It is easy to verify that the system of Eqs. (18) and (35)--(40) has a
nontrivial solution, if and only if $y=2$ and this solution is
\begin{equation}
q^{\pm}(l_{0},l_{1}) = \left\{ \begin{array}{rl}
(-1)^{l_{1}-1} (l_{1}+l_{0}) q^{\pm}_{0} & 
{\mbox{\rm for even}} \hspace{0.3 cm} l_{1}+l_{0},\\
\pm (-1)^{l_{1}-1} (l_{1}+l_{0}) q^{\pm}_{0} & 
{\mbox{\rm for odd}} \hspace{0.3 cm} l_{1}+l_{0},
\end{array} \right.
\end{equation}
where $q^{\pm}_{0}$ are arbitrary constants. 

Therefore, nontrivial solutions for the system of Eqs. (18) and (35)--(40) in
the quantities $q^{\pm}(l_{0},l_{1})$ do not exist in the domain of admissible
values of the parameter $y$ ($y \in (-\infty, 0)$), i.e., there are no nonzero
operators $Q_{\pm}$ in variant 2 of Lagrangian (3). It follows from the
conditions of type (19) that any operator $Q^{\tau lm}_{\pm}$ ($\tau \in 
S^{B}$) is linearly and homogeneously expressed in terms of the operators
$Q_{\pm}$. Consequently, all the operators $Q^{\tau lm}_{\pm}$ in the considered
variant are zero, and interaction Lagrangian (3) is trivial.

\begin{center}
{\large \bf 5. An example of the mass spectrum in the infinite-component field
theory with the double symmetry}
\end{center}

Announcing the completion of the first stage of the construction of the theory
of the infinite-component ISFIR-class fields, we demonstrate some properties of
the mass spectrum of the fermionic states in such a theory with one example
without details of proofs.

Let a free fermionic field theory be assigned the description given in
Corollary 2 (point 2) in [1] with $k_{1}=3/2$, and let the mass term of its
Lagrangian arises from interaction Lagrangian (3) in variant 1A as a result of a
spontaneous secondary-symmetry breaking.

The state vector $\psi_{M}$ of a particle of mass $M$ in its rest frame must
satisfy a relativistically invariant equation (of Gelfand--Yaglom type (I.1))
taking the form
\begin{equation}
(M \Gamma^{0} - R)\psi_{M} = 0
\end{equation}
and the normalization condition
\begin{equation}
|\overline{\psi_{M}} \Gamma^{0} \psi_{M}| < +\infty
\end{equation}
if $M$ is a discrete point of the mass spectrum or
\begin{equation}
\overline{\psi_{M'}} \Gamma^{0} \psi_{M} = a_{0}\delta (M-M')
\end{equation}
($a_{0}$ is some number) if $M$ and $M'$ belong to the continuous part of the
mass spectrum.

We find the operator $\Gamma^{0}$ in Eq. (42) using relations (I.18)--(I.23),
(I.79), and (I.80). We define the operator $R$ by formula (4), where we set
$\kappa = 0$ and use relation (17) for the operator $Q^{(0,1)00}_{+}$. 

Because the operator $\Gamma^{0}$ is diagonal with respect to the spin index 
$l$ and the spin-projection index $m$, the vector states $\psi_{M}$ are 
assigned certain values of spin and its projection. We let $J$ denote the spin 
of the particle in its rest frame. Because, according to condition 4 in [1],
the operators $\Gamma^{0}$ is Hermitian and its matrix elements in the case
considered are real, the state vectors $\psi_{M}$ satisfying Eq. (42) possess
the certain spatial parity $P\psi_{M}=\pm \psi_{M}$, i.e.,  
$(\psi_{M})_{(-1/2,l_{1})Jm} = \pm (\psi_{M})_{(1/2,l_{1})Jm}$. Taking the
abovementioned into account, we write relation (42) as a system of equations in
the components $\psi (l_{1}) \equiv (\psi_{M})_{(1/2,l_{1})Jm}$ of the state
vector with positive parity. We have
$$\left( l_{1}-\frac{1}{2}\right) \sqrt{(l_{1}+J+1)(l_{1}-J)} 
\psi (l_{1}+1) + \left[ (-1)^{l_{1}+\frac{1}{2}} 2 l_{1}\left( J+\frac{1}{2}
\right) \right.$$
\begin{equation}
\left. +\frac{2\lambda}{Mc_{0}} \left( l_{1}^{2}-\frac{1}{4}\right) 
q^{+}(\frac{1}{2},l_{1}) \right] \psi (l_{1}) + \left( l_{1}+
\frac{1}{2}\right) \sqrt{(l_{1}+J)(l_{1}-J-1)} \psi (l_{1}-1)=0,
\end{equation}
where $J \geq l_{1}-1$ and $l_{1} \geq 3/2$. We calculate the quantities
$q^{+}(1/2,l_{1})$ using recursive relations (18) and (28) with $z > 2$.

Using the system of equation (18) and (28), we find the asymptotic behavior as 
$l_{1} \rightarrow +\infty$, 
\begin{equation}
q^{+}(\frac{1}{2},l_{1}) = r_{0}\frac{u^{l_{1}-\frac{1}{2}}}{l_{1}-\frac{1}{2}}
\left( 1 + {\cal O}\left( \frac{1}{l_{1}}\right) \right),
\end{equation}
where $u=(z+ \sqrt{z^{2}-4})/2$ and $r_{0}$ is a constant depend on the
parameter $z$. 

Using Eqs. (45), we next find the asymptotic behavior of the quantity 
$\psi (l_{1})$ as $l_{1} \rightarrow +\infty$. We obtain
$$\psi (l_{1}) = r_{1}\left( -\frac{2\lambda r_{0}}{Mc_{0}}
\right)^{l_{1}-\frac{1}{2}} 
\frac{u^{[\frac{1}{2}(l_{1}-\frac{1}{2})(l_{1}-\frac{3}{2})]}}
{(l_{1}-\frac{3}{2})!} \left( 1 + 
{\cal O}\left( \frac{1}{l_{1}}\right) \right)$$
\begin{equation}
+ r_{2}\left( -\frac{Mc_{0}}{2\lambda r_{0}}
\right)^{l_{1}-\frac{1}{2}} \frac{(l_{1}-\frac{3}{2})!}
{u^{[\frac{1}{2}(l_{1}-\frac{1}{2})(l_{1}-\frac{3}{2})]}}\left( 1 + 
{\cal O}\left( \frac{1}{l_{1}}\right) \right),
\end{equation}
where $r_{1}$ and $r_{2}$ is constants dependent on the quantities 
$Mc_{0}/\lambda$ and $z$ as on parameters.
 
If the constant $r_{1}$ does not vanish for some value of the quantity $M$,
then the corresponding solution of system of equations (45) in the quantities
$\psi (l_{1})$ evidently satisfies neither normalization condition (43) nor
condition (44). If $r_{1}=0$, then the solution of system (45) satisfies
normalization condition (43), i.e., this solution is the state vector of the
particle, and the corresponding quantity $M$ is its mass. Therefore, if the 
mass spectrum in the considered variant of the theory is not empty for some set
of $z$-parameter values from the interval $z \in (2, +\infty)$, then the 
spectrum is discrete for all values of $z$ from this set.

The normalized solutions of system of equations (45) can hardly be expressed in
terms of some known special functions. We therefore use numerical methods that
allow finding any number of the least values of the mass $M$ and of the
initial components $\psi (l_{1})$ of the corresponding state vectors of the
particles. We choose the value and the sign of the parameter 
$\lambda q_{0}^{+}/c_{0}$ such that the least-enegy state has the positive
spatial parity and its mass is equal to unity. For $z=3.0$, the least masses 
$M$ and the corresponding spin and parity $J^{P}$ of the particles then have 
the values presented in Table 1.

\vspace{0.3 cm}
\hspace{13.3 cm} {\bf Table 1}
\begin{center}
\begin{tabular}
{|p{5.8 mm}|p{7.8 mm}|p{7.8 mm}|p{7.8 mm}|p{7.8 mm}|p{7.8 mm}|
p{7.8 mm}|p{7.8 mm}|p{7.8 mm}|p{7.8 mm}|p{7.8 mm}|p{7.8 mm}|p{7.8 mm}|} 
\hline
& & & & & & & & & & & & \\
$J^{P}$ &$\frac{1}{2}^{+}$ &$\frac{1}{2}^{-}$ &$\frac{3}{2}^{+}$ 
&$\frac{3}{2}^{-}$ &$\frac{5}{2}^{+}$ &$\frac{5}{2}^{-}$ 
&$\frac{7}{2}^{+}$ &$\frac{7}{2}^{-}$ &$\frac{9}{2}^{+}$ 
&$\frac{9}{2}^{-}$ &$\frac{11}{2}^{+}$ &$\frac{11}{2}^{-}$\\
& & & & & & & & & & & & \\
\hline
$M$ &1.00 &1.99 &1.59 &3.44 &2.95 &6.71 &5.97 &14.0 &12.7 &30.4 &28.1 &68.0
\\ 
 &4.16 &9.04 &7.68 &17.5 &15.5 &36.5 &33.1 &79.3 &73.2 &178. &166. &406.
\\
 &20.2 &46.1 &40.6 &95.5 &86.6 &207. &191. &464. &433. & & &
\\
 &107. &251. &227. &544. &501. & & & & & & &
\\
 &597. & & & & & & & & & & &
\\
\hline
\end{tabular}
\end{center}

This example allows making the following conclusion related to the constructed
theory of the infinite-component ISFIR-class fields. There exist
relativistically invariant equations of Gelfand--Yaglom type (I.1) such that,
first, their mass spectrum has no continuous part, and, second, each value of
spin and parity is assigned a countable set of masses that is not bounded from
above, and, third, the minimum value of mass for the given spin grows 
infinitely with the spin. These conclusions qualitatively correspond to the 
picture that is provided by the parton model of hadrons together with the
concept of bags for quarks and gluons. A more detailed analysis of the mass
spectra in the constucted infinite-component field theory with the double
symmetry and their relation to the real baryon spectra is the subject of our
subsequent paper.

{\bf Acknowledgments.} The author is grateful to A.U. Klimyk, A.A. Komar, and
V.I. Fushchich for useful discussions of the problems considered in this paper.

\end{document}